
\magnification=1200
\baselineskip=22truept

\def\ff{{\cal F}}

\leftskip=27truepc
\noindent UG--FT--46/94\hfil\break
\noindent July 1995\hfil\break

\rightskip=3.truepc
\leftskip=3.truepc
\vskip 0.2truecm

\centerline{\bf PHYSICAL PARAMETERS AND RENORMALIZATION}
\centerline{\bf OF $U(1)^a\times U(1)^b$ MODELS }
\vskip 0.3truecm

\centerline{F. del Aguila, M. Masip and M. P\'erez-Victoria}
\vskip .5truecm

\centerline{\it Departamento de F\'\i sica
Te\'orica y del Cosmos}
\centerline{\it Universidad de Granada}
\centerline{\it 18071 Granada, Spain}
\vskip 0.7truecm

\centerline{ABSTRACT}
\vskip 0.3truecm
We analize the structure of models with
unbroken and spontaneously broken
$U(1)^a\times
U(1)^b$ gauge symmetry.
We show that the
quantum corrections to the $2N$ gauge charges, with
$N=$ number of fermions $+$ number of scalars,
can be absorbed in the
redefinition of {\it three} independent gauge couplings
($g^a$, $g^b$, and $g^{ab}$).
We establish the (one-loop) conditions
on the matter content for $g^{ab}=0$ (a value
usually assumed in the literature) and we show that in the
minimal extensions of the Standard Model with an extra
$U(1)$ symmetry the choice $g^{ab}=0$ is not stable
under radiative corrections induced by the standard
Higgs fields. Moreover, $g^{ab}=0$ to
all orders seems to require an exact symmetry. The
spontaneous breaking of the gauge symmetry
induces further mixing between the two gauge bosons
and introduces a fourth independent physical
parameter. A consequence of our analysis is that
the usual tree-level description with
only three physical parameters ({\it i.e.}, two
gauge couplings and one gauge boson mixing angle) is not
in general a justified zero order limit of the
treatment including radiative corrections.

\vfill\eject

\leftskip=2.pc
\rightskip=1.pc
\vskip 0.3truecm
\noindent{\bf 1. Introduction.}
\vskip 0.2truecm

The extensions of the Standard Model (SM) with an extra
$U(1)$ gauge symmetry have been extensively
studied during the last years [1].
They appear as a possible low-energy limit in many grand
unified scenarios [2], and they
are not banished to very high
energies by present data [3].
As a matter of fact,
precision experiments at LEP as well as direct
searches at large hadron colliders (TEVATRON) set
(stringent) limits on new gauge interactions,
but do not exclude their discovery at future colliders
(LHC or NLC) [4].

Usually these analyses assume definite models with
few free parameters. In this way, the fits (which often
depend also on few independent observables) are simplified.
Beyond tree level,
however, the number of free parameters is related to
the number of independent renormalized parameters. Hence, if
a parameter is not let to vary, one must
make sure that the constrained model is stable
under quantum corrections.
In the case of extra gauge interactions this is a delicate
point [5]. In this paper we study at the one-loop level models
which include a sector with
$U(1)^a\times U(1)^b$ gauge symmetry. From our
analysis it follows
that a generic extension of the SM with gauge
group $SU(3)_C\times SU(2)_L\times U(1)_Y\times U(1)_{Y'}$
has four new free parameters: the mass $M_{Z'}$ of the
extra vector boson $Z'$; the mixing angle $\phi$ between
the $Z$ (mass eigenstate) and $Z'^0$
($Y'$-current eigenstate)
vector bosons; the overall strength $g_2$ of the new $Y'$
current; and the mixing $g_{12}$ [6] of the $Y'$
current with the standard hypercharge $Y$.
In particular, we prove that $g_{12}$ is
a free parameter in these models:
it is  physical (to be determined experimentally) and
necessary to absorb
the {\it infinities} when calculating quantum corrections.
$g_{12}$ can be consistently ignored if an extra
symmetry is present in the theory but this is not the case
in many popular models.
We do not claim that the effects due to a
non-vanishing $g_{12}$ are always sizeable and they
could not be neglected in a tree-level analysis.
However, it is worth to emphasize that:

$\bullet$ When obtaining experimental bounds on
$Z'$ models, the nonstandard (tree-level) contributions
are often calculated assuming $g_{12}=0$ and varying
the extra $Y'$ charge (for example, considering
different combinations of the two nonstandard
$U(1)$ subalgebras in $E_6$).
It seems more systematic, however, to stick to a
particular
model (which corresponds to a definite $Y'$ charge),
and to let
all its free parameters including $g_{12}$ vary, rather
than constraining the
whole class of models with a particular choice of one free
parameter.

$\bullet$ The three gauge couplings $g_1$, $g_2$, and $g_{12}$
cover the whole parameter space of a model with a
gauge symmetry subgroup $U(1)_Y\times U(1)_{Y'}$.
For example, the  $U(1)_Y\times U(1)_{Y'} \subset
SU(2)_R\times U(1)_{B-L}$ sector of
left-right symmetric models (resulting from $SO(10)$)
is a particular case with the two new gauge couplings
related:
$$
g_1 = {{g_Rg_{B-L}}\over {\sqrt {{2\over 5} g^2_R +
{3\over 5} g^2_{B-L}} }},\
g_2 = \sqrt {{2\over 5} g^2_R +
{3\over 5} g^2_{B-L}} , \
g_{12} = {{g_R^2 - g_{B-L}^2}\over {\sqrt {{5\over 3} g^2_R +
{5\over 2} g^2_{B-L}} }}
\eqno(1)
$$
(this model is known as the $\chi$ model in the
literature (see Ref.~[1] for definitions)).
In general, since
the three parameters get renormalized and {\it run} with the
scale, they bring information of other (larger) scales
(which may point out to grand unification, left-right symmetry
at higher scales, etc). In particular, if $g_1, \ g_2$ and
$g_{12}$ satisfy Eq.~(1) at some scale, it
indicates that there is left-right symmetry restoration
at that scale.

$\bullet$ At any rate, in the absence of extra
symmetries, a fully consistent one-loop analysis
of precision data including a relatively light $Z'$
requires considering $g_{12}$ as a physical parameter.

As a first step to analyse $Z'$
extensions of the SM at one loop,
we study in this paper the structure
of models with $U(1)^a\times U(1)^b$ gauge symmetry [7].
In Section 2 we discuss the tree-level Lagrangian. In
Section 3 we fix the choice of renormalized parameters and
introduce our renormalization (on-shell) scheme.
The one-loop renormalization of
the $U(1)^a\times U(1)^b$ model is worked out in detail,
emphasizing the need of an exact extra symmetry
to guarantee that $g_{12}$ can be neglected to all orders.
(In the Abelian case discussed here we denote this
gauge coupling $g^{ab}$.)
We assume throughout the paper that the theory is vectorlike,
although our examples are based on realistic extensions of
the SM. Thus, it must not create any confusion when we refer
to $SO(10)$ or left-right models to specify  $U(1)^a\times U(1)^b$
matter contents (models). The results which we illustrate
with these examples apply in both cases, except for simple
modifications (factors).
In Section 4 we present
the renormalization of
the model with spontaneously broken symmetry. In this case
three renormalized parameters in the Higgs potential are
replaced by the two heavy gauge
boson masses and their mixing angle.
Section 5 is devoted to conclusions.

\vskip 0.3truecm
\noindent{\bf 2. Classical Lagrangian and physical
parameters.}
\vskip 0.2truecm

The classical Lagrangian for $n$ fermions $f_i$
and $m$ scalars $\phi_i$ with
$U(1)^a\times U(1)^b$ gauge symmetry reads
$${\cal L}^{class}=-{1\over 4}F^T_{\mu\nu}
F^{\mu\nu}+\sum\limits_{i=1}^{n}
\overline f_i (i\not D - m_i) f_i +\sum\limits_{i=1}^{m}
(D_\mu\phi_i)^\dagger(D^\mu\phi_i) - V(\phi_i)\;,\eqno(2)
$$
where the antisymmetric tensor
$$F_{\mu\nu}={F^a_{\mu\nu}\choose F^b_{\mu\nu}}=
{{\partial_\mu A^a_\nu - \partial_\nu A^a_\mu}
\choose {\partial_\mu A^b_\nu - \partial_\nu A^b_\mu}}
\;\eqno(3)
$$
and the covariant derivatives
$$\eqalign{&D_\mu f_i=
 \partial_\mu f_i+ i(\tilde q^a_i\;
\tilde q^b_i) {A^a_\mu \choose A^b_\mu }f_i\;,\cr
&D_\mu \phi_i=
\partial_\mu\phi_i+ i(\tilde Q^a_i\;
\tilde Q^b_i) {A^a_\mu \choose A^b_\mu }\phi_i\;.\cr}
\eqno(4)
$$
$A^{a,b}$ are the two gauge boson fields, and
$\tilde q^{a,b}_i$ and $\tilde Q^{a,b}_i$ the $2N$
($N=n+m$) fermion and scalar charges, respectively,
whereas $V(\phi_i)$ is a polinomial of at most fourth order
preserving the $U(1)^a\times U(1)^b$ gauge symmetry.
${\cal L}$ is the most general Lagrangian renormalizable
by power counting and invariant under the transformations
$$\eqalign{f_i\rightarrow & \exp \{ -i (\tilde q^a_i\;
\tilde q^b_i) {\theta^a\choose \theta^b}\}\; f_i\;;\cr
\phi_i\rightarrow & \exp \{ -i (\tilde Q^a_i\;
\tilde Q^b_i) {\theta^a\choose \theta^b}\}\; \phi_i\;;\cr
{A^a_\mu \choose A^b_\mu }
& \rightarrow {A^a_\mu \choose A^b_\mu } +
{\partial_\mu \theta^a\choose \partial_\mu \theta^b}\;,\cr}
\eqno(5)
$$
with $\;\theta^{a,b}$ the two gauge parameters.

The invariance of $F_{\mu\nu}$ under gauge transformations
also allows for a gauge kinetic term
of the form $\;F^T_{\mu\nu}K F^{\mu\nu}\;$, with $K$ an
arbitrary  $2\times 2$ symmetric matrix. $K$ can be absorbed,
however, into a vector boson field redefinition
(note that a redefinition of $A_\mu$ also redefines
the $2N$ charges of the matter fields in Eq.~(4)).
Without loss of generality we can then assume
$K=-{1\over 4} I$, still leaving the arbitrarity of
rotating the two gauge fields. Then only $2N-1$
charges are physical, since the rotation left, which
is related to the impossibility of distinguishing on
physical grounds between the two degenerate (massless)
gauge bosons,
can be used to fix one of the $2N$ charges to zero.
Hence the $2N$ charges are determined fixing one charge
conventionally and fitting $2N-1$ independent experiments.
We assume fermion fields with masses $m_i$, whereas
$V(\phi_i)$ includes scalar masses and couplings.
We also assume that the Yukawa couplings are forbidden
by some symmetry, for they are not important for
our discussion.

In summary, ${\cal L}$ in Eq.~(1) is a generic (classical)
Lagrangian
of at most dimension four with $U(1)^a\times U(1)^b$
gauge symmetry. This is not altered at the quantum level:
the theory is renormalizable [8] and gauge invariance
does not allow for any other term. Among the physical
parameters of the model, however, quantum corrections
can be used to
distinguish between those which are {\it renormalized}
(and in this sense are free) from those which are
{\it constants}. For example, in QED with just a $U(1)$ gauge
symmetry and $N$ matter fields there is one free parameter,
the electric charge $e$ usually identified with the charge
of the proton, and $N-1$ constants, the ratios of the
remaining charges to $e$. As we shall show, the gauge
sector in a model
with $U(1)^a\times U(1)^b$ symmetry and $N$ matter fields
depends on {\it three} free parameters and $2N-4$
constants: the charges $(\tilde q_i^{a}\;\tilde q_i^{b})$
can be splitted into
$$(\tilde q^a_i\;\tilde q^b_i)\equiv
(q^a_i\;q^b_i)\pmatrix{g^{a}&g^{ab}\cr
0&g^{b}\cr}\;,\eqno(6)
$$
where $(q^a_i\;q^b_i)$ are constant charges
(four of them fixed
arbitrarily) and $g^{a}$, $g^{ab}$, and $g^{b}$
three parameters (gauge couplings)
which will absorb all the quantum corrections.
(We use only one superscript, $a,b$, for diagonal terms
$g^a, g^b$.)
In general, beyond tree level,
just two gauge couplings $g^a$ and $g^b$
(one for each $U(1)$ subgroup)
are not enough to renormalize the theory.

Obviously for $N=1$ there is only one free
parameter (independent charge), since in this case
the gauge fields can be rotated to decouple completely one
gauge boson.
For $N>1$, three experiments involving
two matter fields with independent charges
(let say $\tilde q_{1,2}^{a}\;\tilde q_{1,2}^{b}$)
can be used to fix $g^{a}$, $g^{ab}$ and $g^{b}$ (once
fixed one $\tilde q$ charge and the 4 charges
$(q^a_{1,2}\;q^b_{1,2})$ conventionally);
the remaining charges would then be fixed
after determining $(\tilde q_i^{a}\;\tilde q_i^{b}), \
i = 3, ..., N$ from $2N-4$
independent experiments:
$$
(q^a_i\;q^b_i)=
(\tilde q^a_i\;\tilde q^b_i)\pmatrix{
{1\over g^{a}}&-{g^{ab}\over g^ag^b}\cr
0&{1\over g^{b}}\cr}\;,\;\; (i=3,...,N)\;.\eqno(7)
$$

In spontaneously broken theories the former discussion
applies but the gauge boson mass eigenstate bases are
fixed and there is no freedom to rotate them. Hence,
in the broken case there are $2N$ physical charges and $2N$
independent experiments are needed to fix them. Then
Eq.~(6) remains general, $g'^{ba} \not= 0$ (we use a prime
to denote the couplings to mass eigenstates),
$$
(\tilde q'^a_i\;\tilde q'^b_i)\equiv
(q^a_i\;q^b_i)\pmatrix{g'^{a}&g'^{ab}\cr
g'^{ba}&g'^{b}\cr}\equiv
(q^a_i\;q^b_i)\pmatrix{g^{a}&g^{ab}\cr
0&g^{b}\cr}\pmatrix{\cos \phi&-\sin \phi\cr
\sin \phi&\cos \phi\cr}\;.\eqno(8)
$$
$\phi$ is the angle rotating from the
gauge boson basis triangularizing the
renormalized gauge coupling
matrix to the renormalized gauge boson mass eigenstate
basis.

\vskip 0.3truecm
\noindent{\bf 3. Renormalization of $U(1)^a\times U(1)^b$:
unbroken case.}
\vskip 0.2truecm

In this Section we study the renormalization of a theory with
unbroken gauge symmetry $U(1)^a\times U(1)^b$. We discuss
the parametrization of the gauge couplings (valid to all
orders) and work out in detail their renormalization in
the on-shell scheme at one loop. We show that even the
popular (minimal) extensions of the standard model with one
extra $U(1)$ require two new gauge couplings in order to
cancel the divergent contribution of the Higgs
fields. Moreover, even if the models are enlarged adding
extra matter in order to fulfil the one-loop conditions for
consistently neglecting the second gauge coupling $g^{ab}$,
there is no guarantee for the cancellation of infinities
at two loops. As a matter of fact, in the
examples we have looked at the cancellation of infinities
to all orders requires an exact extra symmetry:
a more general gauge invariance or its discrete
remnant.

The renormalized Lagrangian in terms of bare quantities
has the same expression as the classical Lagrangian
in Eq.~(2)
$$
\eqalign{
{\cal L} = &-{1\over 4}F^{0\;T}_{\mu\nu}
F^{0\mu\nu}
+ \sum\limits_{i=1}^{n}
\overline f^0_i (i\not{D\;} - m^0_i) f^0_i \cr
& +\sum\limits_{i=1}^{m}
[(D_\mu\phi^0_i)^\dagger(D^\mu\phi^0_i) - \mu_i^{0\ 2}
{\phi^0_i}^\dagger\phi^0_i] \cr
& - V^{(>2)}(\phi^0_i) - {1\over 2}
\partial^\mu A^{0\;T}_\mu \xi^{0\ -1}
\partial^\nu A^0_{\nu}\;,\cr}\eqno(9)
$$
where $\mu_i^0$ are the
scalar masses and $V^{(>2)}$ contains the terms of dimension
3 and 4. A covariant gauge fixing term
has been added. In this gauge the ghosts decouple.
Both ultraviolet and infrared divergences are regularized
using dimensional regularization. Renormalized fields
and couplings are related to these bare quantities
(we denote
fermion and scalar charges by little q when referring to
both)
$$
\eqalign{&
A^0_\mu = Z^{1\over 2}_A A_\mu\;;\;\;
f^0_i = Z^{1\over 2}_{f_i} f_i\;;\;\;
\phi^0_i = Z^{1\over 2}_{\phi_i} \phi_i\;;\cr
&\tilde q^{0\;T}_i=
\tilde q^T_iZ_{\tilde q_i}\;;\;\;
m^0_i=m_i+\delta m_i\;;\;\;
\mu^{0\ 2}_i=\mu^2_i+\delta \mu^2_i\;;\cr}
\eqno(10)
$$
and analogously for the couplings in $V^{(>2)}$ and for
$\xi ^{0\ -1}$.
$A^{(0)}_\mu$ and $\tilde q^{(0)}_i$ are 2 dimensional
vectors and $\xi ^{(0)}, Z^{1\over 2}_A$ and
$Z_{\tilde q_i}$ are $2\times 2$ matrices.
The non-diagonal terms generate counterterms which
will be needed to cancel infinities.
The gauge
symmetry translates into Ward identities for
Green functions. In particular for renormalized
one-particle irreducible Green functions,
$$
\partial _z ^{\ \mu} \langle \bar f_i (x) f_i (y)
A_{\mu} ^r (z) \rangle ^{\rm irre} \ = \
i \tilde q_i^u Z_{\tilde q_i}^{uv} Z_A^{{1\over 2}\ vr}
\langle \bar f_i (x) f_i (y) \rangle ^{\rm irre}
(\delta (z-y) - \delta (z-x))
\; , \eqno(11)
$$
and analogously for scalar fields.
The finiteness of the other quantities in Eq.~(11) implies
that the product $Z_{\tilde q_i} Z_A^{{1\over 2}}$
is also finite. As a matter of fact
$$
Z_{\tilde q_i} Z_A^{{1\over 2}} = 1\eqno(12)
$$
in appropriate renormalization schemes such as
minimal subtraction and on-shell (see the Appendix).
Then $Z_{\tilde q_i}$ is independent of the matter
field $i$ and equal to $ Z_A^{-{1\over 2}}$; and Eq.~(11)
for the gauge couplings reads [9]
$$
(\tilde q^{0\ a}_i\;\tilde q^{0\ b}_i) =
(\tilde q^a_i\;\tilde q^b_i)
\pmatrix{Z^{-{1\over 2}\;aa}_{A}&Z^{-{1\over 2}\;ab}_{A}\cr
Z^{-{1\over 2}\;ba}_{A}&Z^{-{1\over 2}\;bb}_{A}\cr}\;.
\eqno(13)
$$
This is the generalization of the constancy to all orders
of the charge ratios in QED to the case of
$U(1)^a\times U(1)^b$. Splitting
$$
(\tilde q^a_i\;\tilde q^b_i) \equiv
(q^a_i\;q^b_i)\pmatrix{g^{a}&g^{ab}\cr
g^{ba}&g^{b}\cr}\; ,
\eqno(14)
$$
and similarly for the bare couplings, Eq.~(13) implies
$$
\eqalign{(q^{0\ a}_i\;q^{0\ b}_i) = &\ (q^a_i\;q^b_i)\;
;\cr
\pmatrix{g^{0\ a}&g^{0\ ab}\cr
g^{0\ ba}&g^{0\ b}\cr} = &\pmatrix{g^{a}&g^{ab}\cr
g^{ba}&g^{b}\cr}
\pmatrix{Z^{-{1\over 2}\;aa}_{A}&Z^{-{1\over 2}\;ab}_{A}\cr
Z^{-{1\over 2}\;ba}_{A}&Z^{-{1\over 2}\;bb}_{A}\cr}
\;.\cr}
\eqno(15)
$$
Hence, it is possible also in this case to define charges
which do not renormalize, $q^{(0)\ a,b}_i$, but to absorb
all quantum corrections we must introduce a $2\times 2$
matrix of gauge couplings,
$\pmatrix{g^{(0)\ a}&g^{(0)\ ab}\cr
g^{(0)\ ba}&g^{(0)\ b}\cr}$. To determine them, 4 charges
defining a $2\times 2$ invertible matrix,
{\it e.g.}, $q^{a,b}_{1,2}$, must be fixed conventionally
in Eq.~(14).
In the unbroken case with $U(1)\times U(1)$ gauge symmetry,
however, Eq.~(15) is too general.
The freedom to define (rotate) the renormalized gauge
fields in Eq.~(10) allows to assume $g^{ba} = 0$ in
Eqs.~(14) and (15), and thus the matrix $g$ triangular.
Besides, the freedom to rotate the gauge bosons in Eq.~(9)
allows to assume $Z^{{1\over 2}\;ba}_A = 0$ in Eq.~(10),
and thus the matrix $Z^{1\over 2}_A$ (and its inverse
$Z^{-{1\over 2}}_A$ in Eqs.~(13) and (15)) triangular.
Both minimal subtraction and on-shell schemes are
compatible with this choice.
Under this
rotation the gauge fixing matrix $\xi ^{0\ -1}$
also transforms, but it was arbitrary, although fixed.
(The Ward identity for the gauge boson propagator implies
that the gauge fixing term does not renormalize,
{\it i. e.},
$\xi ^{0\ -1} = Z^{-{1\over 2}\;T}_{A} \xi ^{-1}
Z^{-{1\over 2}}_{A}$). With the former choices the
right-hand side of Eq.~(15) gives a triangular
$g^0$ matrix:
$$
\pmatrix{g^{0\ a} & g^{0\ ab}\cr
0 & g^{0\ b}\cr} = \pmatrix{g^{a} & g^{ab}\cr
0 & g^{b}\cr}
\pmatrix{Z^{-{1\over 2}\;aa}_{A} & Z^{-{1\over 2}\;ab}_{A}\cr
0 & Z^{-{1\over 2}\;bb}_{A}\cr}\;.
\eqno(16)
$$
This is our main result: the renormalization of the
gauge couplings in models with two abelian gauge symmetries
requires three couplings
$g^{(0)\ a}, g^{(0)\ ab}, g^{(0)\ b}$, satisfying Eq.~(16).
We have used the freedom existing
in defining the degenerate (massless) gauge bosons.

Let us make explicit this analysis to one loop.
Following the on-shell scheme prescription in the Appendix
we evaluate the renormalized vector boson proper
selfenergies.
These can be written as the sum
of transverse and longitudinal parts:
$$i\Pi_{\mu\nu}^{rs}(q^2)=
i[{\cal A}^{rs}(q^2)(g_{\mu\nu}-
{q_\mu q_\nu\over q^2})+{\cal B}^{rs}(q^2)
{q_\mu q_\nu\over q^2}]\;.\eqno(17)
$$
Using the Feynman
rules in Ref.~[10] with $\xi = I$, we find from the
diagrams in Fig.~1 (excluding the fourth diagram
which only contributes in the broken case)
$$
\eqalign{\pmatrix{{\cal A}^{aa}&{\cal A}^{ab}\cr
{\cal A}^{ba}&{\cal A}^{bb}\cr}&=
-q^2 {1\over 16\pi^2}\lbrace \; \sum\limits_{i=1}^{n}\lbrack
{4\over 3} C_{UV}-8\int_{0}^{1}\!dx\; x(1-x)
\ln D_{f_i} \rbrack\pmatrix{(\tilde q^a_i)^2&
\tilde q^a_i\tilde q^b_i\cr
\tilde q^b_i\tilde q^a_i&(\tilde q^b_i)^2\cr}\cr
 + \sum\limits_{i=1}^{m}&\lbrack
{1\over 3} C_{UV}+{1\over 3}+{2\over q^2}(
\int_{0}^{1}\!dx\; D_{\phi_i} \ln D_{\phi_i}  -
\mu^2_i\ln \mu^2_i)\rbrack \pmatrix{(\tilde Q^a_i)^2&
\tilde Q^a_i\tilde Q^b_i\cr
\tilde Q^b_i\tilde Q^a_i&(\tilde Q^b_i)^2\cr}\rbrace\cr
-q^2 &\pmatrix{2({Z^{{1\over 2}\ aa}_A}-1)&
{Z^{{1\over 2}\ ba}_A}+
{Z^{{1\over 2}\ ab}_A}\cr
{Z^{{1\over 2}\ ab}_A}+{Z^{{1\over 2}\ ba}_A}&
2({Z^{{1\over 2}\ bb}_A}-1)\cr}\;,\cr}
\eqno(18)
$$
with
$C_{UV}=({1\over \epsilon}-\gamma+\ln 4\pi)$, $\epsilon=
(4-d)$, $D_{f_i}=m_i^2- q^2x(1-x)$, and
$D_{\phi_i}=\mu_i^2- q^2x(1-x)$.
The last term in Eq.~(18) stands for the one-loop
counterterms. They result from expanding
$Z_A = Z_A^{{1\over 2}\ T} Z_A^{{1\over 2}}$:
$$
\eqalign{
\pmatrix{ \sum\limits_{r=a,b} Z^{{1\over 2}\ ra}_{A}
Z^{{1\over 2}\ ra}_{A} &
\sum\limits_{r=a,b} Z^{{1\over 2}\ ra}_{A}
Z^{{1\over 2}\ rb}_{A} \cr
\sum\limits_{r=a,b} Z^{{1\over 2}\ rb}_{A}
Z^{{1\over 2}\ ra}_{A} &
\sum\limits_{r=a,b} Z^{{1\over 2}\ rb}_{A}
Z^{{1\over 2}\ rb}_{A} \cr}
 = &
\pmatrix{1&0\cr
0&1\cr} + \cr
 & \pmatrix{2({Z^{{1\over 2}\ aa}_A}-1)&
{Z^{{1\over 2}\ ba}_A}+
{Z^{{1\over 2}\ ab}_A}\cr
{Z^{{1\over 2}\ ab}_A}+{Z^{{1\over 2}\ ba}_A}&
2({Z^{{1\over 2}\ bb}_A}-1)\cr} \; + \; ...\; . \cr }
\eqno(19)
$$
Eq.~(19) makes apparent that ${\cal A}^{rs}$
(which is symmetric) in (18) and the corresponding on-shell
conditions (gauge invariance assures ${\cal A}^{rs}(0) = 0$)
$$
{\partial {\cal A}^{rs}\over \partial q^2}\vert_{q^2= 0}=0\;,
\eqno(20)
$$
are independent of the choice (rotation) of
the bare gauge field basis.
At one loop this means that the three conditions in Eq.~(20)
fix $Z^{{1\over 2}\; aa}_A,\ Z^{{1\over 2}\; bb}_A$, and
$Z^{{1\over 2}\; ab}_A + Z^{{1\over 2}\; ba}_A$.
Hence,
we can assume in agreement with Eq.~(16) that
$Z^{{1\over 2}\; ba}_A = 0$ and use Eq.~(20) to fix
the remaining $Z^{1\over 2}_A$ matrix elements,
and in turn the matrix elements of the inverse matrix
$Z^{-{1\over 2}}_A$:
$$
\eqalign{
Z^{-{1\over 2}\ aa}_A & = 1 +
{1\over 32\pi^2}\lbrace \; \sum\limits_{i=1}^{n}
{4\over 3} \lbrack C_{UV}-
\ln m_i^2 \rbrack (\tilde q^a_i)^2
+\sum\limits_{i=1}^{m}
{1\over 3} \lbrack C_{UV}-
\ln \mu^2_i)\rbrack (\tilde Q^a_i)^2 \rbrace \; ,\cr
Z^{-{1\over 2}\ ab}_A & =
{1\over 16\pi^2}\lbrace \; \sum\limits_{i=1}^{n}
{4\over 3} \lbrack C_{UV}-
\ln m_i^2 \rbrack \tilde q^a_i \tilde q^b_i
+\sum\limits_{i=1}^{m}
{1\over 3} \lbrack C_{UV}-
\ln \mu^2_i)\rbrack \tilde Q^a_i \tilde Q^b_i \rbrace \; ,\cr
Z^{-{1\over 2}\ bb}_A & = 1 +
{1\over 32\pi^2}\lbrace \; \sum\limits_{i=1}^{n}
{4\over 3} \lbrack C_{UV}-
\ln m_i^2 \rbrack (\tilde q^b_i)^2
+\sum\limits_{i=1}^{m}
{1\over 3} \lbrack C_{UV}-
\ln \mu^2_i)\rbrack (\tilde Q^b_i)^2 \rbrace \; .\cr
}
\eqno(21)
$$
Thus, in general $Z^{-{1\over 2}\ ab}_A$ is infinite and
then $g^{0\ ab}$ too (see Eq.~(16)).

The corresponding on-shell
conditions on the fermion and the scalar
selfenergies fix the field renormalization constants
$Z^{1\over 2}_{f_i}$ and $Z^{1\over 2}_{\phi_i}$ and
the mass countertems $\delta m_i$ and $\delta \mu_i$,
whereas the scalar three- and four-point
functions are renormalized by the appropiate choice of
renormalization constants.

It is interesting to know under which assumptions one
can neglect $g^{ab}$, because it is convenient
to have as few free parameters as possible
when performing fits to experimental data.
At any rate many existing bounds
on gauge extensions of the standard model have been
obtained fixing $g^{ab} = 0$.
($g^{ab}$ is a physical parameter and its experimental
value can be compatible with zero accidentally.)
The question is whether it
renormalizes or not. Generically the answer depends on
the renormalization scheme. What we really want to know
is if a scheme (and a model) exists where
$Z^{-{1\over 2}\ ab}_A = 0$.
In this case $g^{ab}/g^b$ is constant under renormalization
and the particular choice $g^{ab} = 0$ is stable and consistent
(although not necessary).
At one loop the infinite part of $Z^{-{1\over 2}\ ab}_A$
cancels if (assuming $g^{ab} = 0$)
$$
{4\over 3}\sum\limits_{i=1}^{n} q^a_i q^b_i +
{1\over 3}\sum\limits_{i=1}^{m} Q^a_i Q^b_i = 0
\eqno(22)
$$
(see Eq.~(21)). In chiral theories
${4\over3}$ is replaced by
${2\over3}$ and $i$ runs over
the 2-component spinors.
Eq. (22) is fulfilled if
the fermion and scalar fields define {\it complete} multiplets
of a simple group containing one (or both)
$U(1)$ factor(s). For example, this is the case if the matter
contents of the $U(1)^a\times U(1)^b$ model defines
complete multiplets of $SU(2)^a\times U(1)^b(\supset
U(1)^a\times U(1)^b)$ or $SO(10)(\supset U(1)^a\times U(1)^b)$.
We do not see, however, any necessity
(based on anomaly cancellation, minimality,
or grand unification) to assume this, specially in
the scalar sector. In particular, consider
the minimal model in Table 1 where $SO(10)$ is
broken at very large scales ($\approx 10^{15}$ GeV) to
$SU(3)_C\times SU(2)_L\times U(1)^a\times U(1)^b$, with
$U(1)^a$ the hypercharge, $Y$, and $U(1)^b$ the extra $U(1)$
in $SO(10)$, usually denoted $U(1)_{\chi}$ [1]. If one
assumes a minimal fermion content of 3
chiral families in the
{\bf 16} representation, all of them {\it survive} the
breaking of $SO(10)$ and the fermion contribution to
$g^{Y\chi}$ is zero
($\sum\limits_{i=1}^{n} q^Y_i q^{\chi}_i = 0$
for the fermions in Table 1).
In the scalar sector, however,
one usually
accommodates the Higgs doublets in the
{\bf 10} representation of $SO(10)$; when this
group is broken there is no reason to keep the
leptoquarks in the {\bf 10} light, with masses identical
to those of the Higgs fields (on the contrary, it is
phenomenologically preferred to give them large masses).
The same argument applies in supersymmetric extensions
of the SM with an extra $U(1)$. Then radiative corrections
induced by the light Higgses
generate a nonzero $g^{Y\chi}$ gauge coupling even in
the minimal scenarios.
$\sum\limits_{i=1}^{m} Q^Y_i Q^{\chi}_i = 2 {1\over 5}
{\sqrt {3\over 2}}$ for the
scalars in Table 1: one $SU(2)_L$ doublet and one singlet,
together with their complex conjugated representations.
One can insist
in adding extra light matter (scalars) in order to satisfy
Eq.~(22) but this would not guarantee that
$Z^{-{1\over 2}\ ab}_A$ is finite at two loops. For
instance, the first diagram in Fig. 1 with a gauge boson
crossing the fermion bubble vertically is proportional to
$\sum\limits_{x=a,b} \sum\limits_{i=1}^{n} q^a_i q^x_i
q^b_i q^x_i$ (and similarly for other diagrams). In contrast
with the corresponding one-loop contribution this two-loop
diagram is infinite for the $\chi $ model
$$
\sum\limits_{x=Y, \chi} \sum\limits_{i=1}^{n} q^Y_i q^x_i
q^{\chi}_i q^x_i = 3 (-{1\over {10}} {\sqrt {1\over 6}})
.\eqno(23)
$$
And there is no reason for cancellations among diagrams.
Hence one expects (although small) nonzero contributions
to $g^{Y\chi}$ after renormalizing from the unification
scale [9].

Let us compare this model with the $LR$ model [1]
in the same Table:
$U(1)^a$ is the third component of $SU(2)_R$, $T_3^R$, and
$U(1)^b$ is the baryon minus the lepton ($B-L$) number,
$Q_{B-L}$.
This model is
also contained in $SO(10)$ and if we also assume a minimal
fermion contents of 3 chiral families in the {\bf 16}
representation, $\sum\limits_{i=1}^{n} q^R_i q^{B-L}_i = 0$
at one loop and
$\sum\limits_{x=R, B-L} \sum\limits_{i=1}^{n} q^R_i q^x_i
q^{B-L}_i q^x_i = 0$ at two loops. In fact
$Z^{-{1\over 2}\ RB-L}_A = 0$ to all orders. This follows
from the vanishing of ${\cal A}^{RB-L}$ in Eq.~(18), what
is guaranteed by an
exact symmetry interchanging $u^c \leftrightarrow d^c,
e^c \leftrightarrow {\nu}^c$ and changing the sign of the
$R$ gauge boson, $A^R \rightarrow -A^R$ but leaving
unchanged the $B-L$ gauge boson, $A^{B-L}$. The Higgs
sector has to be enlarged to maintain ${\cal A}^{RB-L} = 0$:
at least one scalar must be added with the same quantum
numbers as $e^c$ in order to complete an $SU(2)_R$ doublet
with $N^c$, and similarly for $\overline N^c$.

These two models illustrate the different cases:

$\bullet$ In general $Z^{-{1\over 2}\ ab}_A$ is infinite and
$g^{ab}$ is not only a physical parameter but a necessary
one to absorb the infinities of the theory, as in the
$U(1)_Y\times U(1)_{\chi}$ model.

$\bullet$ If as in the $LR$ model $U(1)_R\times U(1)_{B-L}$
there is an exact symmetry
requiring $Z^{-{1\over 2}\ ab}_A = 0$, $g^{ab}$ can be
consistently neglected.
The exact symmetry in this model is a discrete
remnant of the $SU(2)_R$ symmetry embedded in $SO(10)$.

$\bullet$ If a $U(1)$ factor is part of a non-abelian
gauge group then gauge invariance guarantees the
vanishing of $Z^{-{1\over 2}\ ab}_A$ and
$g^{(0)\ ab}$. In the $LR$ model this is guaranteed by
$SU(2)_R\times U(1)_{B-L}$. (This is similar to
the SM case where $Z^{-{1\over 2}\ LY}_A = 0$ is implied by
$SU(2)_L\times U(1)_Y$.)

$\bullet$ In the $\chi $ model one can gauge $SU(5)$
(which contains  the hypercharge), completing matter and
vector boson representations, to guarantee
$Z^{-{1\over 2}\ Y\chi}_A$ and $g^{(0)\ Y\chi}$ zero.
Then
$\sum\limits_{x\in SU(5)\times U(1)_{\chi}}
\sum\limits_{i=1}^{n} q^Y_i q^x_i q^{\chi}_i q^x_i = 0$.

It is worth to emphasize that although
$Y = {\sqrt {3\over 5}} T_3^R + {\sqrt {2\over 5}} Q_{B-L}$,
$Q_{\chi} =
{\sqrt {2\over 5}} T_3^R - {\sqrt {3\over 5}} Q_{B-L}$,
the (generalized) $\chi$ and $LR$ models are equivalent
only if $g^{ab}$ is included:

$\bullet$ $g^{RB-L} \not= 0$ violates the (discrete)
symmetry and $Z^{-{1\over 2}\ RB-L}_A$ is infinite.

$\bullet$ If we $write$ the $LR$ model ($g^{RB-L} = 0$)
in the $Y, \chi$ basis, $g^Y,\ g^{\chi}$ and $g^{Y\chi}$
(as well as $Z^{-{1\over 2}\ Y,\chi ,Y\chi}_A$) are
related (see Eq.~(1)).

$Z^{-{1\over 2}}_A$ in the on-shell scheme (Eq.~(21))
has also finite contributions.
(In the minimal substraction scheme
there are no such contributions.) They also cancel if there
is an exact symmmetry distinguishing $a$ and $b$ and
constraining the fermion and scalar masses, as in the
$LR$ model. If the masses violate the symmetry, one must
expect that they will generate infinite $Z^{-{1\over 2}\ ab}_A$
contributions at higher orders, and a nonzero $g^{ab}$.

\vskip 0.3truecm
\noindent{\bf 4. Renormalization of
spontaneously broken $U(1)^a\times U(1)^b$.}
\vskip 0.2truecm

The results of the unbroken case apply to the
spontaneously
broken phase [8]. It will be more convenient, however,
to make a different choice of gauge fixing term
in order to simplify real calculations and of
renormalization conditions to improve the comparison
with data in extended electroweak models.

In the broken phase the scalar fields in
Eq.~(9) with nonvanishing
VEVs $v^0_i,\ i = 1, ..., l$,
(that we assume to be real)
$$
\phi^0_i={1\over \sqrt{2}}(v^0_i+h^0_i+i\chi^0_i)\;,
\eqno(24)
$$
are shifted.
The term in the Lagrangian involving the covariant
derivative of these
scalars gives rise to the vector boson mass matrix:
$$
\eqalign{  \sum\limits_{i=1}^{l}
(D_\mu\phi^0_i)^\dagger(D^\mu\phi^0_i)
 = & \sum\limits_{i=1}^{l}\lbrace
{1\over 2}(\partial_\mu h^0_i\partial^\mu h^0_i +
\partial_\mu \chi^0_i\partial^\mu \chi^0_i)
+\chi^0_i\partial_\mu h^0_i \tilde Q^{0\ T}_i A^{0\ \mu} + \cr
(v^0_i+h^0_i)\partial_\mu \chi^0_i
\tilde Q^{0\ T}_i & A^{0\ \mu} +
{1\over 2}A^{0\ T}_\mu \tilde Q^{0}_i
\tilde Q^{0\ T}_i A^{0\ \mu}
(2v^0_ih^0_i+h^0_ih^0_i+\chi^0_i\chi^0_i)\rbrace  + \cr
& {1\over 2} A^{0\ T}_\mu M^{0\ 2} A^{0\ \mu}, \cr }
\eqno(25)
$$
with $\tilde Q^{0}_i = \pmatrix{g^{0\ a} & 0 \cr
g^{0\ ab} & g^{0\ b}\cr} \pmatrix{Q^{0\ a}_i\cr
Q^{0\ b}_i \cr}$ and
$M^{0\ 2} = \sum\limits_{i=1}^{l} \tilde Q^{0}_i v_i^{0\ 2}
\tilde Q^{0\ T}_i$.
This can be diagonalized rotating the gauge boson basis,
$M^{0\ 2} = R^T_{\phi^0} M^{0\ 2}_d R_{\phi^0}$, where
$$
M^{0\ 2}_d = \pmatrix{M_a^{0\ 2}&0\cr
0&M_b^{0\ 2}\cr},\
R_{\phi^0} = \pmatrix{\cos \phi^0&\sin \phi^0\cr
- \sin \phi^0&\cos \phi^0\cr} \;,
\eqno(26)
$$
and $\phi ^0$ is the angle defining the rotation from the
basis where the gauge coupling matrix takes a triangular
form (Eq.~(16)) to the mass eigenstate basis,
$A'^0_{\mu} = R_{\phi^0} A^0_{\mu}$. (Prime refers to
gauge boson mass eigenstates.)
In this basis
$$
\pmatrix{g'^{0\ a}&g'^{0\ ab}\cr
g'^{0\ ba}&g'^{0\ b}\cr} = \pmatrix{g^{0\ a}&g^{0\ ab}\cr
0&g^{0\ b}\cr} R^T_{\phi^0}\;.
\eqno(27)
$$

In order to simplify real computations
we work in a $R_{\xi}$ gauge [11] where
the vector-scalar mixing in Eq.~(25)
cancels
$$
{\cal L}_{GF}^0 = -{1\over 2} \ff^{0\ T} \xi^{0\ -1}
\ff^0\;,\eqno(28)
$$
with $\xi ^{0\ -1}$ a symmetric $2\times 2$ matrix of
gauge parameters and
$$
\ff^0 = \partial_\mu  A^{0\ \mu} - \xi^0
\sum\limits_{i=1}^{l}\tilde Q^0_i v^0_i \chi^0_i \;.
\eqno(29)
$$
We must also add to the Lagrangian the corresponding
Faddeev-Popov term for the ghosts
${c^{0\ a}\choose c^{0\ b}}$,
which in this gauge do not decouple,
$$
{\cal L}^0_{ghost}=
-\overline c^{0\ T} [\partial^{\alpha}\partial_{\alpha} +
\xi^0 \sum\limits_{i=1}^{l} \tilde Q^0_i v^0_i
( v^0_i + h^0_i ) \tilde Q^{0\ T}_i]c^0\; .
\eqno(30)
$$

For electroweak precision tests it may be adequate to
choose as free parameters the gauge boson masses,
what motivates to use the on-shell scheme (see the Appendix).
The set of independent parameters in the unbroken case,
$2N-1$ gauge couplings, $\tilde q_i^{0\ r}$, the fermion
masses and the scalar couplings (masses), is replaced
in the broken case by the 2 gauge boson masses,
$M_r^{0\ 2}$, $2N$ gauge couplings $\tilde q'^{0\ r}_i$,
the fermion masses and the same but 3 scalar couplings
(masses). This means trading three scalar couplings (masses)
by the three
parameters fixing the (symmetric) gauge boson mass matrix,
the two mass eigenvalues and the rotation angle $\phi ^0$
in Eq.~(26).
This angle is included in the gauge coupling
definition (see Eqs.~(8) and (27))
$$
(\tilde q'^{0\ a}_i\ \tilde q'^{0\ b}_i) =
(\tilde q^{0\ a}_i\ \tilde q^{0\ b}_i)\ R^T_{\phi^0} =
(q^{0\ a}_i\ q^{0\ b}_i)
\pmatrix{g^{0\ a}&g^{0\ ab}\cr
0&g^{0\ b}\cr}
\pmatrix{\cos \phi^0&-\sin \phi^0\cr
\sin \phi^0&\cos \phi^0\cr}\;.
\eqno(31)
$$
In the symmetric case there are $2N-1$ independent charges
because the gauge boson basis is defined up to a rotation,
which is fixed conventionally. In the broken case the
gauge boson basis (mass eigenstates) is fixed and the
$2N$ charges (including the rotation angle) will be
determined fitting $2N$ independent experiments.
Two remarks are in order. In some models not
all the parameters are independent. Since the
scalar charges $\tilde Q_i^{0\ r}$ defining the
mass matrix $M^{0\ 2}$ in Eq.~(25) are free parameters,
a general gauge boson mass matrix requires that at least
three different (non-equivalent) scalars get a VEV
($l \geq 3$).
(In order to break both $U(1)$'s, $l$ must be $\geq 2$.)
Thus in the minimal $\chi$ model in Table 1, with
$h$ and $N^c$ only, $\phi ^0$ is a function of the gauge boson
masses [12]. Other parameters in the Higgs potential can be
also replaced by some (of the remaining) VEVs (up to $l-3$).
Otherwise, the VEVs $v^0_i$, which are determined minimizing
the effective potential, are not independent parameters.
They are considered as so, however, when fixing the
corresponding counterterms to satisfy the vanishing
tadpole conditions. All the other counterterms can be
found following the standard procedure [13].

Introducing
as before (Eq.~(10)) renormalized fields and couplings (we
concentrate on the vector boson parameters and fields)
$$
A'^0_\mu = Z^{1\over 2}_{A'} A'_\mu\;;\;\;
\tilde q'^{0\;T}_i=
\tilde q'^T_iZ_{\tilde q'_i}\;;\;\;
M^{0\ 2}_r = M^2_r + \delta M^2_r \;,\;\;
\eqno(32)
$$
the theory can be multiplicatively renormalized. The
Ward identity analogous to that in Eq.~(11), but now
involving also ghosts, implies that the product
$Z_{\tilde q'_i} Z_{A'}^{{1\over 2}}$
is finite. Moreover, the on-shell scheme can be
defined requiring
$Z_{\tilde q'_i} Z_{A'}^{{1\over 2}} = 1$.
Then, Eq.~(13) also applies for $\tilde q'_i$,
$$
\tilde q'^0_i = \tilde q'_i Z^{-{1\over 2}}_{A'}\ ,
\eqno(33)
$$
and splitting the $2N$ charges as in Eq.~(31), we
obtain
$$
\eqalign{(q^{0\ a}_i\;q^{0\ b}_i) = \ & (q^a_i\;q^b_i)\;
{\rm and}\;
\pmatrix{g^{0\ a}&g^{0\ ab}\cr
0&g^{0\ b}\cr}
\pmatrix{\cos \phi^0&-\sin \phi^0\cr
\sin \phi^0&\cos \phi^0\cr} = \cr
\pmatrix{g^{a}&g^{ab}\cr
0&g^{b}\cr} &
\pmatrix{\cos \phi&-\sin \phi\cr
\sin \phi&\cos \phi\cr}
\pmatrix{Z^{-{1\over 2}\;aa}_{A'}&Z^{-{1\over 2}\;ab}_{A'}\cr
Z^{-{1\over 2}\;ba}_{A'}&Z^{-{1\over 2}\;bb}_{A'}\cr}
\;, \cr}\eqno(34)
$$
for the broken case, too.
This equation gives the renormalization of the gauge couplings
and of the gauge mixing angle.

The one-loop expressions for the counterterms are obtained
as before from ${\cal A}'^{rs}$ in Eq.~(17).
The vector boson proper selfenergies,
$i\Pi'^{\alpha\beta}_{\mu\nu}(q^2)$,
receive contributions from the diagrams in Fig.~1.
(See Ref. [10] for the corresponding Feynman rules
in the t'Hooft-Feynman gauge, $\xi = I$.)
The divergent contribution to ${\cal A}'^{rs}$ coming from
the first three diagrams is
the same as in the unbroken case
but exchanging $(\tilde q^a_i\;\tilde q^b_i)$
and $(\tilde Q^a_i\;\tilde Q^b_i)$ in Eq.~(18) by
the corresponding prime charges in Eq.~(8).
The fourth diagram contribution and the wave function
and mass counterterms can be written:
$$
\eqalign{&{\cal A}'_{4+5}=
- {1\over 4\pi^2}
{\sum\limits_{i=1}^{l} \sum\limits_{r=a,b}
\lbrack C_{UV}-\int_{0}^{1}\!dx\;
\ln D_i^r \rbrack
\pmatrix{(\tilde Q'^a_i)^2&\tilde Q'^a_i\tilde Q'^b_i\cr
\tilde Q'^b_i\tilde Q'^a_i&(\tilde Q'^b_i)^2\cr}
(\tilde Q'^r_i)^2 v_i^2} \cr
-& \pmatrix{
2(Z^{{1\over 2}\ aa}_{A'}-1)(q^2-M^2_a)-\delta M^2_a&
Z^{{1\over 2}\ ba}_{A'}(q^2-M^2_b)+
Z^{{1\over 2}\ ab}_{A'}(q^2-M^2_a)\cr
Z^{{1\over 2}\ ab}_{A'}(q^2-M^2_a)+Z^{{1\over 2}\ ba}_{A'}
(q^2-M^2_b)& 2(Z^{{1\over 2}\ bb}_{A'}-1)(q^2-M^2_b)-
\delta M^2_b\cr}\;,\cr}
\eqno(35)
$$
where $D^r_{i}=\mu ^2_ix+M^2_r(1-x)-q^2x(1-x)$.
In the broken
phase the six on-shell conditions on ${\cal A}'^{rs}$,
$$
\eqalign{
&{\cal A}'^{aa}(M_a^2)={\cal A}'^{ab}(M_a^2)=0\;,
\;\;{\partial {\cal A}'^{aa}\over
\partial q^2}\vert_{q^2=M_a^2}=0\;;\cr
&{\cal A}'^{bb}(M_b^2)={\cal A}'^{ab}(M_b^2)=0\;,
\;\;{\partial {\cal A}'^{bb}\over
\partial q^2}\vert_{q^2=M_b^2}=0\;;\cr}
\eqno(36)
$$
fix the counterterms
$Z^{{1\over 2}\ rs}_{A'}$,
$\delta M_r^2$. We find
(we write as in the unbroken case the
$Z^{-{1\over 2}}_{A'}$ renormalization constants)
$$
\eqalign{
Z^{-{1\over 2}\ aa}_{A'}=&\ 1+{C_{UV}\over 32\pi^2}
\lbrace \sum\limits_{i=1}^{n}
{4\over 3} (\tilde q'^a_i)^2
+\sum\limits_{i=1}^{m} {1\over 3}
(\tilde Q'^a_i)^2\rbrace + {\rm finite\ terms}\ ; \cr
Z^{-{1\over 2}\ ab}_{A'}=&\ { C_{UV}\over 16\pi^2}
{1\over M_b^2-M_a^2}\lbrace M_b^2 \lbrack
\sum\limits_{i=1}^{n}
{4\over 3} \tilde q'^a_i\tilde q'^b_i
+\sum\limits_{i=1}^{m}{1\over 3}
\tilde Q'^a_i\tilde Q'^b_i
\rbrack  \cr
& + 4 \sum\limits_{i=1}^{l}
\sum\limits_{r=a,b} v_i^2 (\tilde Q'^r_i)^2
\tilde Q'^a_i\tilde Q'^b_i
\rbrace + {\rm finite\ terms}\ ;\cr
\delta M_a^2=&\ {C_{UV}\over 16\pi^2}
\lbrace M_a^2 \lbrack \sum\limits_{i=1}^{n}
{4\over 3} (\tilde q'^a_i)^2
+\sum\limits_{i=1}^{m}{1\over 3} (\tilde Q'^a_i)^2
\rbrack  \cr
& + 4 \sum\limits_{i=1}^{l}
\sum\limits_{r=a,b} v_i^2 (\tilde Q'^r_i)^2
(\tilde Q'^a_i)^2
\rbrace
+ {\rm finite\ terms}\
.\cr}\eqno(37)
$$
The renormalization constants
$Z^{-{1\over 2}\ ba}_{A'}$,
$Z^{-{1\over 2}\ bb}_{A'}$ and $\delta M_b^2$ can be obtained
from those in Eq~(37)
interchanging the indices $a\leftrightarrow b$.
Note that in the spontaneously broken case there is no
arbitrarity left in the gauge boson basis definition
and then no ambiguity in the determination of the four
independent elements of
the $Z^{-{1\over 2}}_{A'}$ matrix.
These four universal counterterms
are absorbed in the $2\times 2$ gauge coupling matrix
$g'=g R_\phi$. The mixing angle
$\phi$ provides a fourth independente coupling.
All Green functions are then finite.

For later use we chose the on-shell renormalization
scheme. All our results (except for the finite one-loop
contributions to $Z^{-{1\over 2}}_{A'}$) also apply
in the minimal subtraction scheme, which is simpler.

\vskip 0.3truecm
\noindent{\bf 5. Summary and conclusions.}
\vskip 0.2truecm

The extensions of the SM with an extra $U(1)$ symmetry
are a possibility frecuently considered in the literature.
The object of many of these analyses has been to estimate
the (small) effects caused by this new physics on
electroweak observables ($\rho$ parameter, $Z$ width, ...),
and then to use precision data to constrain the
independent parameters
of the models (namely, the mass of the extra neutral
boson $Z'$ and its mixing with the standard gauge boson).
Usually, the way to proceed has been to
combine the SM predictions at one loop with
the nonstandard effects estimated at tree level.
In this framework, the
aim of our study is to discuss the generic procedure
to include (in the on-shell scheme) the
full radiative corrections
in models with two abelian gauge symmetries
$U(1)^a\times U(1)^b$. This is not only
important for consistency but for practical
(numerical) reasons if the extra $Z'$ is
relatively light. We have
proved that to absorb the {\it infinites} of the
theory one needs {\it three} universal gauge couplings
($g^a$, $g^b$, and $g^{ab}$).
In the general case these
couplings are independent parameters to be fixed
experimentally.

If grand unification
at large scales is assumed (for example into
$SO(10)$), then the gauge symmetry implies
$g^{ab}=0$ at the unification scale.
Once the unified symmetry is broken,
however, the order by order conditions on the matter
fields to guarantee $g^{ab} = 0$ may not be satisfied.
This is the case of the usual minimal models
already at one loop, due to the Higgs (scalar) contributions.
At two loops the fermion contributions do not fulfil
the matter conditions either. Then radiative corrections
generate a nonzero $g^{ab}$ at low energies.
In each particular model, this
parameter can be estimated via the renormalization-group
equations.
The $LR$ model is one exception because there is
a discrete symmetry left, reminiscence of the left-right
gauge symmetry, maintaining unmixed the two $U(1)$'s.
$g^{RB-L} \not= 0$ violates (explicitly) this symmetry.
Hence, we conclude that a complete analysis
of $Z'$ effects on precision electroweak data must
contain the third gauge coupling $g^{ab}$. This is
necessary not only to absorb the divergences of the
theory when calculating
beyond the tree level, but also because the low-energy
renormalized value suggested by minimal
unified scenarios (once the leading-log
contributions are taken into account) is different
from zero.

If $g^{ab} \not= 0$ all models with $U(1)\times U(1)$
charges which are linear combination of the
$U(1)^a\times U(1)^b$ charges are equivalent [6].

In the broken case the mixing between the new and the
standard gauge bosons, $\phi$, redefines the currents
and thus the gauge couplings. $g^{ab}$ and $\phi$ are
two independent parameters to be determined measuring
the gauge boson currents (Eq.~(8)).

\vskip 0.3truecm
\noindent{\bf Acknowledgements.}  This work was partially
supported by CICYT under contract AEN94-0936,
by the Junta de Andaluc\'\i a and by the European Union
under contract CHRX-CT92-0004. We thank M. B\"ohm and
W. Hollik for clarifying discussions.

\vskip 0.3truecm
\noindent{\bf Appendix.}
\vskip 0.2truecm

In this Appendix we establish the renormalization
conditions in the complete on-shell scheme [13].

{\it Two-point functions}:

\noindent
{\it Massive case}.
The vector-vector propagator can be separated into
transverse and longitudinal parts:
$$\Delta^{rs}_{\mu\nu}\equiv -i[T^{rs}(q^2)(g_{\mu\nu}-
{q_\mu q_\nu\over q^2})+L^{rs}(q^2)
{q_\mu q_\nu\over q^2}]\;.\eqno(A1)$$
The on-shell conditions are fixed in such a way that
when $\epsilon\equiv (q^2-M^2_a)\rightarrow 0$
the transverse part has a pole in $T^{aa}$ and the other
components are regular (we neglect the
finite width of the particles and consider only the
real part of the propagators):
$$(T^{rs})\vert_{q^2\rightarrow M_a^2}=
\pmatrix{{1\over \epsilon}&O(1)\cr
O(1)&O(1)\cr}\ .\eqno(A2)$$
The selfenergies are corrections to the inverse
propagator. This can be written
$$\Delta^{-1\ rs}_{\mu\nu}\equiv i[T^{-1\ rs}(q^2)(g_{\mu\nu}-
{q_\mu q_\nu\over q^2})+L^{-1\ rs}(q^2)
{q_\mu q_\nu\over q^2}]\;,\eqno(A3)$$
and for $q^2$ close to $M^2_a$
$$(T^{-1\ rs})\vert_{q^2\rightarrow M_a^2}=
\pmatrix{\epsilon + O (\epsilon ^2)&O(\epsilon)\cr
O(\epsilon)&O(1)\cr}\ .\eqno(A4)$$
The behaviour for $q^2\rightarrow M^2_b$ is analogous.
The on-shell conditions on the inverse propagators read
($T^{-1}$ is symmetric)
$$\eqalign{
T^{-1\ aa}(M_a^2)&=T^{-1\ ab}(M_a^2)=0\;,\;\;
{\partial T^{-1\ aa}\over \partial q^2}
\vert_{q^2= M_a^2}=1\;,\cr
T^{-1\ bb}(M_b^2)&=T^{-1\ ab}(M_b^2)=0\;,\;\;
{\partial T^{-1\ bb}\over \partial q^2}
\vert_{q^2= M_b^2}=1\;.\cr}\eqno(A5)$$
Thus, the poles of the transverse gauge boson propagator
coincide with the renormalized gauge boson masses
and the propagator expressions at the poles are the
asymptotic ones.
The corresponding conditions on the selfenergies are
given in the text (Eq.~(35)).

\noindent
{\it Massless case}.
If $M_a^2=M_b^2=0$, on-shell $T^{rs}$ means
$$(T^{rs})\vert_{q^2\rightarrow 0}=
\pmatrix{{1\over \epsilon}&O(1)\cr
O(1)&{1\over \epsilon}\cr}\ ,\eqno(A6)$$
which implies
$$(T^{-1\ rs})\vert_{q^2\rightarrow 0}=
\pmatrix{\epsilon+ O (\epsilon ^2)&O(\epsilon^2)\cr
O(\epsilon^2)&\epsilon+ O (\epsilon ^2)\cr}\ ,\eqno(A7)$$
or
$$\eqalign{
&T^{-1\ aa}(0)=T^{-1\ ab}(0)=
T^{-1\ bb}(0)=0\;,\cr
&{\partial T^{-1\ aa}\over \partial q^2}\vert_{q^2= 0}=
{\partial T^{-1\ bb}\over \partial q^2}\vert_{q^2= 0}=1\;,\cr
&{\partial T^{-1\ ab}\over \partial q^2}\vert_{q^2= 0}=
0\;.
\cr}\eqno(A8)$$
The conditions for the selfenergies are given in the text
(Eq.~(20)). The first three equations in Eq.~(A8) and the
corresponding conditions for the selfenergies are
guaranteed by gauge invariance. The complete set of
conditions is invariant under (renormalized) vector
boson field rotations due to the gauge boson mass
degeneracy. As a result the gauge field counterterms
are fixed up to a rotation, which must be fixed
conventionally (see the text).

Note that the on-shell conditions are imposed only on
$T^{-1\ rs}(q^2)$. Since the theory is renormalizable,
however, the counterterms for $L^{-1\ rs}(q^2)$
(already
fixed by the above conditions on $T^{-1\ rs}(q^2)$)
will cancel the longitudinal divergences and
make finite the full renormalized
inverse propagator. Another comment concerns the
mixing between vector fields with Nambu-Goldstone scalars
$\chi_i$ in models where
the symmetry is spontaneously broken. Since the external
vector fields satisfy the physical polarization
condition $\partial_\mu V_r^\mu = 0$, the vector-scalar
part of the propagator does not contribute to the
$S$-matrix elements and does not affect the
on-shell conditions.

{\it Three-point functions}:

\noindent
{\it Unbroken case}.
The on-shell conditions on
the three-point functions
$-i(\Gamma^{a\ \mu}_i\;\Gamma^{b\ \mu}_i)$ (writing as a
two component row vector the vertices of the two vector
bosons) are
$$(\Gamma^{a\ \mu}_i\;\Gamma^{b\ \mu}_i)
\vert_{\buildrel \not p_1=\not p_2=m_i\over {q^\mu=0}}
=\gamma^\mu(\tilde q^a_i\;
\tilde q^b_i)\;.\eqno(A9)$$

The charge renormalization constants
$Z_{\tilde q_i}$ are fixed by the Ward identity
in Eq. (11) (in momentum space)
$$q_{\mu} \Gamma^{r\ \mu}_i (p, q) =
\tilde q_i^u Z_{\tilde q_i}^{uv} Z_A^{{1\over 2}\ vr}
(S^{-1}_i(p + q) - S^{-1}_i(p))
.\eqno(A10)$$
Differentiating with respect to $q_{\nu}$, setting
$q^\mu=0$ and the external particles on-shell, and using
(A9) and the on-shell conditions on the inverse fermion
propagators
$$
{\partial S^{-1}_i (k)\over \partial {\not k}}
\vert_{\not k = m_i} = 1
,\eqno(A11)$$
we obtain the equality among renormalization constants
$$
Z_{\tilde q_i}=Z_A^{-{1\over 2}}.\eqno(A12)$$

{\it Spontaneously broken case}.
(A9) or the same condition with $q^2=M^2_{a,b}$ lead to
nonuniversal couplings, with corrections depending on
the mass ratios $m^2_i/M^2_{a,b}$. However the corresponding
Ward identities still guarantee
$Z_{\tilde q'_i} Z^{{1\over 2}}_{A'}$ finite.
Thus we can and do impose the renormalization condition
$Z_{\tilde q'_i}=Z^{-{1\over 2}}_{A'}$ order by order.
Although it seems somewhat artificial, this condition
ensures that the radiative corrections are absorbed
into the four universal couplings $g'$.
Note that
the three-point vertices for
on-shell fields will have finite corrections
depending on the fermion masses,
$$\Gamma _i^{r\ \mu}\vert_{\buildrel \not p_1 =
\not p_2=m_i\over {q^2=M^2_r}} = \gamma ^\mu q^s_i g'^{sr} +
{\rm finite\ higher\ order\ terms} \;.\eqno(A13)$$

\vskip 0.3truecm
\noindent{\bf References.}

\noindent [1]
F. Zwirner, Int. J. Mod. Phys. {\bf A3} (1988) 49;
J.L. Hewett and T.G. Rizzo, Phys. Rep. {\bf 183} (1989)193;
P. Langacker and M. Luo, Phys. Rev. {\bf D44} (1991) 817.

\noindent [2]
P. Langacker, R.W. Robinett and J.L. Rosner, Phys. Rev.
{\bf D30} (1984) 1470; P. Candelas {\it et al.}, Nucl. Phys.
{\bf B258} (1985) 46.

\noindent [3]
Review of Particle Properties, Phys. Rev. {\bf D50}
(1994) 1372.

\noindent [4]
A. Djouadi {\it et al.}, Z. Phys. {\bf C56} (1992) 289;
F. del Aguila and M. Cvetic, Phys. Rev. {\bf D50}
(1994) 3158; A. Leike, Z. Phys. {\bf C62} (1994) 265;
D. Choudhury, F. Cuypers and A. Leike, Phys. Lett.
{\bf B333} (1994) 531;
M. Cvetic and S. Godfrey, Univ. of Pennsylvania
preprint, UPR-648-T (hep-ph/9504216).

\noindent [5]
G. Degrassi and A. Sirlin, Phys. Rev. {\bf D40}
(1989) 3066; F. Ey$\beta$elein, Ph. D. Thesis,
W\"urzburg University, 1991.

\noindent [6]
F. del Aguila {\it et al.}, Nucl. Phys. {\bf B272}
(1986) 413;
F. del Aguila, Acta Physica Polonica
{\bf 25} (1994) 1317;
F. del Aguila, M. Cvetic and P. Langacker,
Univ. of Pennsylvania preprint,
UPR-636-T, to appear in Phys. Rev. {\bf D} (hep-ph/9501390).

\noindent [7] The SM extension was considered in Ref. [5];
and it is under study by W. Hollik, E.J. Vargas and F. del
Aguila, M. Masip, M. P\'erez-Victoria.

\noindent [8]
G. 't Hooft, Nucl. Phys. {\bf B35} (1971) 167;
B.W. Lee and J. Zinn-Justin, Phys. Rev. {\bf D5}
(1972) 3121; {\bf D5} (1972) 3137; {\bf D5} (1972) 3155;
B.W. Lee, Phys. Rev. {\bf D9} (1974) 933.

\noindent [9] F.~del Aguila, G.D. Coughlan, and
M. Quir\'os, Nucl. Phys. {\bf B307}
(1988) 633; E {\bf B312} (1989) 751.
This Ref. contains a general discussion of the gauge
coupling renormalization with several $U(1)$ factors,
studying a particular case at two loops. Two remarks
related to the present work are in order. The claim
that the rotation defining the multiplicatively
renormalizable basis is scale independent
at all orders if the renormalization
group trajectory crosses a unification point
is not general. The derivation does not consider
the fact that the second derivative of the
commutator of gauge couplings and $\beta$-functions
does not cancel if the commutator of the
$\beta$-functions and their derivatives is nonvanishing.
In the particular case studied this rotation is scale
independent due to the special matter content.

\noindent [10]
T.P. Cheng and L.F. Li, {\it Gauge theory of elementary
particle physics}, Oxford University Press, 1983.

\noindent [11]
K. Fujikawa, B.W. Lee and A. Sanda, Phys. Rev. {\bf D6}
(1972) 2923;
B.W. Lee and J. Zinn-Justin, Phys. Rev. {\bf D7}
(1973) 1049;
T. Appelquist {\it et al.}, Phys. Rev. {\bf D8}
(1973) 1747.

\noindent [12]
M. Masip and M. P\'erez-Victoria, in preparation.

\noindent [13]
K. Aoki {\it et al.},
Suppl. Progr. Theor. Phys. {\bf 73} (1982) 1;
A. Denner, Fortschr. Phys. {\bf 41} (1993) 4.

\vskip 0.3truecm
\noindent{\bf Table Captions.}

\noindent {\bf Table I:} Charges for the fermions
in the {\bf 16} representation of $SO(10)$ (upper-half):
third component of $SU(2)_L$, $T_3^L$,
normalized hypercharge, $Y$, and extra $\chi$
charge, $Q_{\chi}$. The third component of $SU(2)_R$,
$T_3^R = {\sqrt {3\over 5}} Y + {\sqrt {2\over 5}} Q_{\chi}$
and the baryon minus the lepton number $Q_{B-L} =
{\sqrt {2\over 5}} Y - {\sqrt {3\over 5}} Q_{\chi}$
The bottom-half corresponds to the
minimal Higgs contents in order to break both $U(1)$'s.
Both $h, N$
and $h', \bar N$ are needed in the supersymmetric
case. $h$ and $h'$ complete a {\bf (2,2)} representation
of $SU(2)_L\times SU(2)_R$. $N^c$ and its complex conjugated
$\overline N^c$ are incomplete $SU(2)_R$ representations.

\vskip 0.3truecm
\noindent{\bf Figure Captions.}

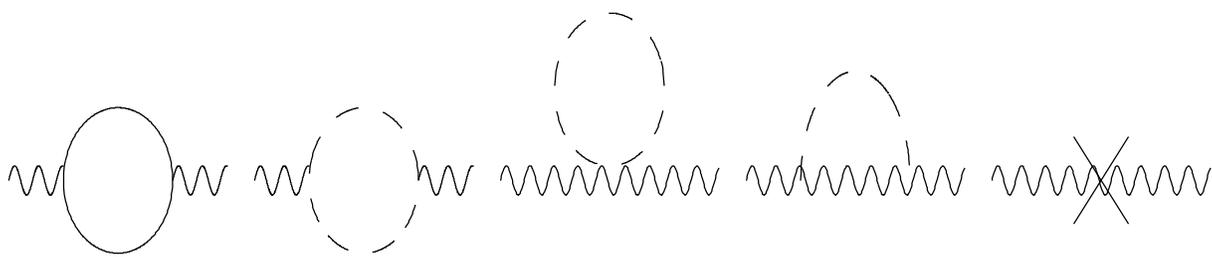
\noindent {\bf Figure 1:} Diagrams contributing
at one loop to the
vector boson selfenergies.
The fourth diagram only contributes in the broken case.

\vfill\eject

$$\vbox{
\tabskip=0.9truecm
\halign{#\hfil&\hfil#\hfil&\hfil#\hfil&\hfil#\hfil&\hfil#\hfil
&\hfil#\hfil&\hfil#\hfil&\hfil#\hfil\cr
\noalign{\hrule}
\omit&\omit&\omit&\omit&\omit&\omit&\omit&\omit\cr
\noalign{\hrule}
\omit&\omit&\omit&\omit&\omit&\omit&\omit&\omit\cr
matter&&$T^L_3$&$Y/{1\over 2}\sqrt{3\over 5}$&
$Q_{\chi}/{1\over 2\sqrt{10}}$&&$T^R_3$&$Q_{B-L}
/{1\over 2}\sqrt{3\over 2}$\cr
\omit&\omit&\omit&\omit&\omit&\omit&\omit&\omit\cr
\noalign{\hrule}
\omit&\omit&\omit&\omit&\omit&\omit&\omit&\omit\cr
$u$&&${1\over 2}$&${1\over 3}$&$-1$&&$0$&${1\over 3}$\cr
$d$&&$-{1\over 2}$&${1\over 3}$&$-1$&&$0$&${1\over 3}$\cr
$u^c$&&$0$&$-{4\over 3}$&$-1$&&$-{1\over 2}$&$-{1\over 3}$\cr
$d^c$&&$0$&${2\over 3}$&$3$&&${1\over 2}$&$-{1\over 3}$\cr
$\nu$&&${1\over 2}$&$-1$&$3$&&$0$&$-1$\cr
$e$&&$-{1\over 2}$&$-1$&$3$&&$0$&$-1$\cr
$e^c$&&$0$&$2$&$-1$&&${1\over 2}$&$1$\cr
$\nu^c$&&$0$&$0$&$-5$&&$-{1\over 2}$&$1$\cr
\omit&\omit&\omit&\omit&\omit&\omit&\omit&\omit\cr
\noalign{\hrule}
\omit&\omit&\omit&\omit&\omit&\omit&\omit&\omit\cr
$h^+$&&${1\over 2}$&$1$&$2$&&${1\over 2}$&$0$\cr
$h^0$&&$-{1\over 2}$&$1$&$2$&&${1\over 2}$&$0$\cr
$N^c$&&$0$&$0$&$-5$&&$-{1\over 2}$&$1$\cr
\omit&\omit&\omit&\omit&\omit&\omit&\omit&\omit\cr
\noalign{\hrule}
\omit&\omit&\omit&\omit&\omit&\omit&\omit&\omit\cr
$h'^0$&&${1\over 2}$&$-1$&$-2$&&$-{1\over 2}$&$0$\cr
$h'^-$&&$-{1\over 2}$&$-1$&$-2$&&$-{1\over 2}$&$0$\cr
$\overline N^c$&&$0$&$0$&$5$&&${1\over 2}$&$-1$\cr
\omit&\omit&\omit&\omit&\omit&\omit&\omit&\omit\cr
\noalign{\hrule}
\omit&\omit&\omit&\omit&\omit&\omit&\omit&\omit\cr
\noalign{\hrule}
\omit&\omit&\omit&\omit&\omit&\omit&\omit&\omit\cr
}}$$
\centerline{{\bf Table I}}
\vskip 0.3truecm

\vfill\eject\end